\documentclass[acmlarge]{acmart}

\setcopyright{none}

\makeatletter
\newcommand{\myconfshort}{\acmConference@shortname}
\newcommand{\myconffull}{\acmConference@name}
\newcommand{\myconfdate}{\acmConference@date}
\newcommand{\myconfloc}{\acmConference@venue}
\AtBeginDocument{
  \fancypagestyle{firstpagestyle}{
    \fancyhead{}\fancyfoot[C]{}}
  \fancyhf{}
  \fancyhead[LO]{\@headfootfont\shorttitle}\fancyhead[RE]{\@headfootfont\@shortauthors}\fancyhead[LE]{\@headfootfont\footnotesize \myconfshort, \myconfdate, \myconfloc}\fancyhead[RO]{\@headfootfont\footnotesize \myconfshort, \myconfdate, \myconfloc}\fancyfoot[C]{}}
\makeatother
\acmBooktitle{\conffull\@ (\confshort), \confdate, \confloc}

\usepackage{subcaption}
\usepackage{url}

\usepackage{float}  \usepackage{placeins}  \usepackage{stfloats}

\usepackage{moreverb,url}

\usepackage{xcolor}
\usepackage{tabularx}
\usepackage[normalem]{ulem}

\usepackage{booktabs}
\usepackage[table,xcdraw]{xcolor}
\usepackage{makecell}
\usepackage{tabularx}

\usepackage{tcolorbox}
\usepackage{enumitem}
\tcbuselibrary{listingsutf8} 

\usepackage{subcaption}
\usepackage{graphicx}

\usepackage{comment}
\usepackage{etoolbox}
\usepackage{xcolor}
 
\usepackage{graphicx} 
\usepackage{float}

\title[Factors that Lead to Non-Development or Abandonment of AI Systems]{To Build or Not to Build? Factors that Lead to Non-Development or Abandonment of AI Systems}

\author{Shreya Chappidi}
\affiliation{\institution{University of Cambridge}
  \city{Cambridge}
  \country{United Kingdom}
}
\affiliation{\institution{National Cancer Institute, National Institutes of Health}
  \city{Bethesda, MD}
  \country{United States}
}

\author{Jatinder Singh}
\email{jat@compacctsys.net}
\affiliation{\institution{Research Centre Trust, UA Ruhr, University Duisburg-Essen}
  \city{Duisburg}
  \country{Germany}
}
\affiliation{\institution{University of Cambridge}
  \city{Cambridge}
  \country{United Kingdom}}
\date{}

\copyrightyear{2026}
\acmYear{2026}
\setcopyright{usgovmixed}
\acmConference[FAccT '26]{The 2026 ACM Conference on Fairness, Accountability, and Transparency}{June 25--28, 2026}{Montreal, QC, Canada}
\acmBooktitle{The 2026 ACM Conference on Fairness, Accountability, and Transparency (FAccT '26), June 25--28, 2026, Montreal, QC, Canada}
\acmDOI{10.1145/3805689.3806736}
\acmISBN{979-8-4007-2596-8/2026/06}

\begin{document}

\begin{abstract}

Responsible AI research typically focuses on examining the use and impacts of deployed AI systems. 
Yet, there is currently limited visibility into the pre-deployment decisions to
pursue building such systems in the first place.
Decisions taken in the earlier stages of development shape which systems are ultimately released, and therefore represent potential, but underexplored, points for intervention. 
As such, this paper investigates factors influencing AI non-development and abandonment throughout the development lifecycle. Specifically, we 
first perform a scoping review of academic literature, civil society resources, and grey literature including journalism and industry reports. 
Through thematic analysis of these sources, we develop a taxonomy of six categories of factors contributing to AI abandonment: ethical concerns, stakeholder feedback,  development lifecycle challenges, organizational dynamics, resource constraints, and legal/regulatory concerns. 
Then, we collect data on real-world cases of AI system abandonment via an AI incident database and a practitioner survey to evidence and compare factors that drive abandonment both prior to and following system deployment. 
While academic responsible AI communities often emphasize ethical risks as reasons to not develop AI, our empirical analysis of these cases demonstrates the diverse, and often non-ethics-related, levers that motivate organizations to abandon AI development. 
Synthesizing evidence from our taxonomy and related case study analyses, we identify gaps and opportunities in current responsible AI research to (1) engage with the diverse range of levers that influence organizations to abandon AI development, and (2) better support appropriate (dis)engagement with AI system development.

\end{abstract}



\makeatletter
\let\@authorsaddresses\@empty
\makeatother

\maketitle

\section{Introduction}

There are growing calls and concerns over responsible AI (RAI) development and deployment, across academic, industry, legal/policy, and public spheres. 
Many academic works interrogate and propose artifacts to improve development practices, including auditing datasets and their creation \cite{gebru_datasheets_2021},
toolkits for fairness evaluations \cite{lee_landscape_2021}, and approaches to improve model transparency  \cite{mitchell_model_2019}. 
Emerging regulation
seeks to govern algorithms through protections against automated decision-making and scrutiny including impact assessments for high-risk AI systems \cite{hollanek_toolkit_2025, euaiact}. 
There are also growing civil advocacy concerns over increasing AI 
development,
including journalism and associated databases documenting harms from deployed systems \cite{raji_fallacy_2022, aiaaic_ai_2025}
and general skepticism regarding AI hype and overuse \cite{stop_gen_ai_stop_2025}. 

These calls and tools for supporting RAI development often advocate for critical evaluation practices and the appropriate governance mechanisms. 
At the same time, many critical decisions are made earlier in the AI development lifecycle---including explicit or implicit decisions to pursue AI development in the first place---that are less frequently explored in RAI works \cite{kawakami_responsible_2024,greene_better_2019}. 
Existing analyses of AI disuse or abandonment largely focus on systems that were deployed and subsequently monitored (either formally or informally) \cite{johnson_fall_2024}. 
As such, works addressing AI development challenges often focus on mitigating risks that are surfaced, or could surface, post-deployment (even if the resulting tool or method itself affects pre-deployment stages, such as dataset bias for example) \cite{kawakami_responsible_2024, wong_seeing_2023, greene_better_2019}.

In contrast, there is limited empirical research investigating {challenges}  in AI development that occur \textit{before} deployment, particularly in scenarios where AI development is abandoned or not pursued at all, creating multiple gaps in RAI research. 
First, it is critical to interrogate \textit{all} blockers to AI development, as they can reveal early-stage failure points 
in development processes that later stage interventions may not appropriately or effectively address (e.g., fairness evaluations may not {always} mitigate {or reveal issues that stem from decisions made} during early-stage problem formulation) \cite{passi_problem_2019, wang_against_2024, cobbe_reviewable_2021}. 
Second, investigating AI abandonment can reveal real-world incentives that shape whether and how AI systems are built. 
For example, examining what kinds of AI systems are abandoned and for what reasons can highlight what criteria are considered and prioritized during organizational decision-making.  
A deeper understanding of these broad and diverse incentives can enable the RAI community to better orient and target processes to support practitioners during AI development. 
Lastly, non-development is also generally an underexplored decision that relates to AI governance. As such, investigating abandoned AI development cases 
can make non-development more visible and defensible as an RAI practice, and reveal levers that influence decision-making to (not) pursue development. 
By focusing on cases where AI development is abandoned or not pursued, we seek to identify current gaps in tooling, processes, and incentives that shape AI (non-)development in practice.

\subsubsection*{Definitions.}
{In this paper, we define \emph{\textbf{non-development}} as organizational decisions to \textit{not} build a proposed system, and \textbf{\emph{abandonment}} as organizational decisions to \textit{stop} building or using an AI system. AI abandonment can occur at any time, including prior to deployment, such as during early-stage data collection or late-stage testing and validation during the AI development lifecycle, or following deployment.}
{We also note that AI abandonment is viewed through a context-dependent lens, and not inherently negative or positive.
In some cases, abandonment can be beneficial, preventing development of nonfunctional, unethical, or underresourced AI systems. 
Conversely, abandonment may be undesirable for well-designed systems that could offer tangible benefits, but face practical challenges leading to or after deployment.}

\subsection{Contributions}
To characterize, analyze, \& increase visibility of factors leading to AI abandonment\slash non-development, this paper:
\begin{enumerate}
    {\item performs a scoping review of academic and grey literature to surface and \textbf{establishes a taxonomy of factors contributing to AI non-development and abandonment}, organizing the factors into six categories: ethical concerns, resource constraints, development lifecycle challenges, legal\slash regulatory concerns, stakeholder feedback, and organizational dynamics;}

    {\item {empirically analyzes real-world cases of AI abandonment} collected via a public AI incident database and an online practitioner survey, \textbf{providing evidence on how both ethical concerns and a diverse range of other factors drive AI abandonment and non-development in real-world contexts};}
\item \textbf{highlights gaps and opportunities in current responsible AI research, tooling, and practice} to {investigate and engage levers influencing AI non-development and abandonment.}
\end{enumerate}

\section{Related Work}

AI systems research often focuses on model\slash system performance (including benchmarks, evaluations, and fairness metrics) \cite{eriksson_can_2025, fodor_line_2025}, or 
the \textit{adoption} {or impacts of using} systems that have already been built (and often deployed).
However, given broad critiques of automation and other risks posed by AI systems, it is critical to explore conditions that lead to disengagement with technologies, from development to deployment.

\subsubsection*{AI (non-)adoption.}
Many human-computer interaction (HCI) studies \cite{kelly_what_2023, russo_navigating_2024, choung_trust_2023} draw upon the technology acceptance model (TAM) \cite{davis_perceived_1989, venkatesh_technology_2008} to explore end user adoption {(including uptake and subsequent patterns of use)}  of AI systems. 
{For example, \citet{russo_navigating_2024} examines TAM, diffusion of innovation theory \cite{rogers_diffusion_1983}, and social cognitive theory to 
understand why employees adopt generative AI, arguing that adoption is driven by compatibility of tools with existing workflows rather than conventional TAM factors like usefulness.}
Meanwhile, other industrial-organizational psychology works find that performance expectations, organizational resources \cite{veiga_ethics_2024}, cultural values \cite{oshaughnessy_what_2021} and social dynamics of transferring knowledge \cite{wijnhoven_organizational_2022} drive successful AI adoption.

These works offer complex sociotechnical models mediating AI adoption and use, but mainly concern systems that have already been built and often focus on user-specific factors like trust. 
While adoption studies provide critical insights into {whether deployed AI systems will be adopted and used}, they can overlook broader factors mediating the development process itself, including initial decisions to even pursue AI development. Thus, to address this gap, we now {discuss indicative} literature on \textit{not} pursuing AI development.

{\subsubsection*{AI non-functionality or non-use.}
Other works have 
proposed non-use of algorithms under specific conditions. 
\citet{raji_fallacy_2022} create a taxonomy of AI `functionality' issues, arguing for critical reflection over what AI systems can actually do (and not do) and recommend legal, policy, and organizational interventions to improve functional safety of AI systems. 
Their primary focus on failures relating to the AI system itself can importantly support identification of (in)appropriate AI use cases, but they do not specifically discuss non-development or focus on the ``infrastructural or environmental `meta' failures'' \cite{raji_fallacy_2022} that may influence decisions to abandon AI development. 
Focusing on AI nonuse, \citet{bhatt_when_2024} propose algorithmic resignation as a form of governance
where deployed algorithms could be intentionally decommissioned under certain conditions.

\subsubsection*{AI non-development}

While prior work has examined AI non-adoption at the deployment stage and AI non-use in specific scenarios, there is relatively less work focusing on the conditions that encourage or discourage AI development in the first place. 
Some academic works \cite{baumer_when_2011} and civil tech initiatives \cite{jordan_dont_2021} urge practitioners to refrain from building unnecessary or objectionable systems. 
Meanwhile, \citet{mun_particip-ai_2024} explicitly examine lay perceptions over the hypothetical impacts of \textit{not} developing AI tools, finding preferences to avoid techno-solutionism but also fears over lost potential benefits from not developing certain use cases.
Yet, there has been limited focus on the principle of `not building' in the growing context of real-world AI development.

Two key studies offer empirical insights into AI non-development. First, \citet{kawakami_studying_2024} use semi-structured interviews and design activities to explore how U.S. public sector agencies decide whether to create or adopt new AI tools. They highlight pressure from external organizations as critical influences on public sector decision-making, indicating the need to explore these factors across broader contexts and application domains. 
Next, \citet{johnson_fall_2024} investigate 40 public cases of AI abandonment, outlining a common six-stage process of ``campaigns to abandon an algorithm.'' The work provides important foundations for investigating decomissioned AI systems; however, it conceptualizes abandonment as a reactive process concerned over the ability of an algorithm to cause harm. 
As such, the work primarily analyzes cases of post-deployment AI abandonment due to real or potential harms, and does not explore other reasons why AI development may be abandoned (e.g., resource constraints, etc.). 
While \citet{johnson_fall_2024} offer a process to specifically describe \textit{how} organizations abandon AI systems, there remains a gap in establishing the factors \textit{why} organizations abandon AI development, particularly where systems are not deployed or lack clear potential for harm.

\subsubsection*{Summary.}

Further research is needed to investigate why and how why individuals, teams, or organizations choose to pursue AI development, and also how development processes might be challenged pre-deployment. Both cases of choosing to \textit{not} develop or \textit{stop} developing can result in systems that never get deployed, rendering these cases underexplored in RAI literature. Our work aims towards this gap, investigating factors mediating decisions to abandon AI development across diverse domains and industries.

\section{A taxonomy of factors contributing to AI abandonment}
\label{sec:taxonomy}
We establish a taxonomy of factors contributing to AI abandonment to improve understanding of both 1) decision-making processes to pursue AI development and 2) diverse failure points, challenges, and trade-offs that can lead to abandonment. We developed our taxonomy through a scoping multivocal literature review \cite{garousi_guidelines_2019} that includes both academic and grey literature sources \cite{lu_responsible_2024}, including civil society organizations and resources, and technology journalism. The high-level goal of this research was to identify \textit{What factors contribute to AI system non-development or abandonment?}

\subsubsection*{Methodology}
\label{sec:taxonomy-methodology}
To build the taxonomy, we perform a multivocal scoping literature review, including both academic and grey literature, {to explore influences on} why AI development may be abandoned or not pursued. We incorporate \textit{grey literature}, including civil society resources, technology journalism, and industry reports and white papers, to capture other timely, pragmatic factors that could drive organizational decision-making and may be overlooked in traditional academic databases \cite{garousi_guidelines_2019, kamei_grey_2021, garousi_benefitting_2020}.

\begin{figure*}[hb]
    \centering
    \includegraphics[width=1\linewidth]{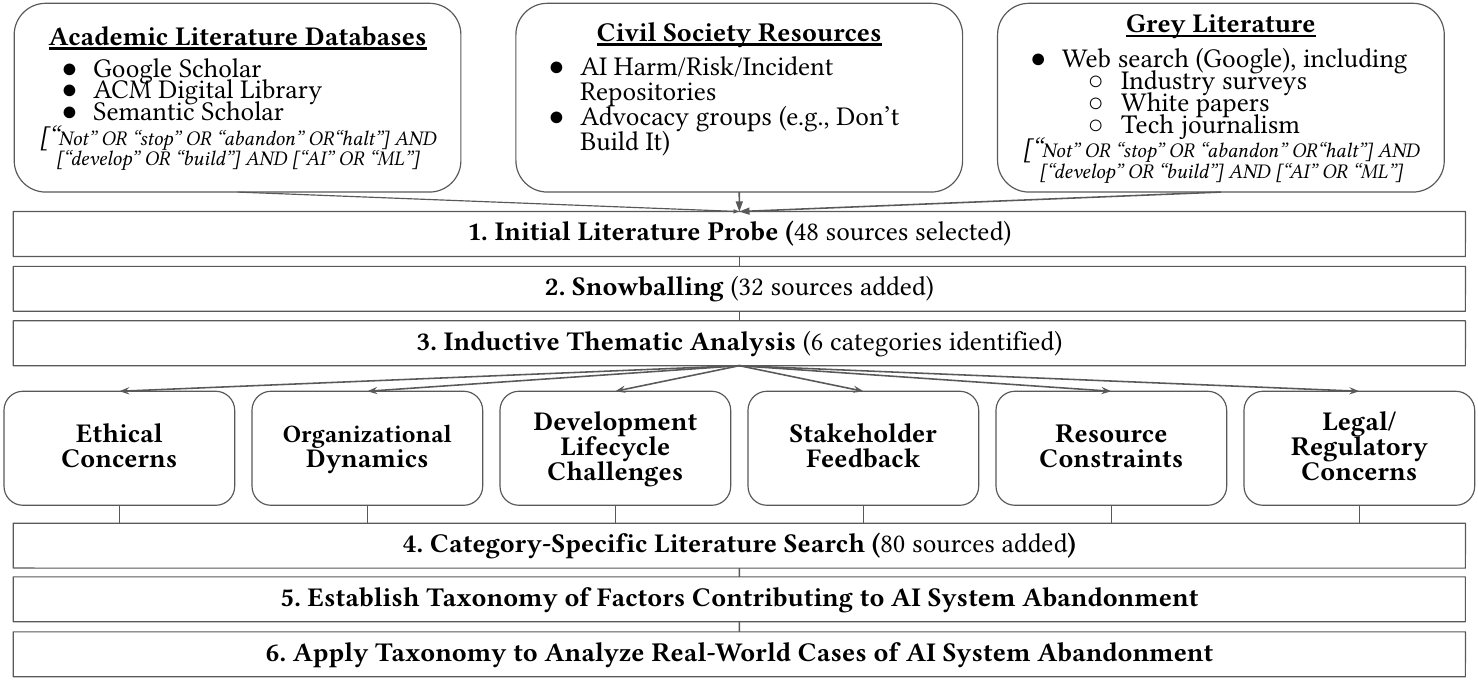}
    \caption{{Overall methodology, including scoping literature workflow, used to establish and analyze categories of factors contributing to AI abandonment.}}
\label{fig:taxonomy-methods}
\end{figure*}

Using search terms and sources detailed in Figure \ref{fig:taxonomy-methods}, we performed preliminary queries across academic literature databases and grey literature sources (web search) for relevant publications {discussing factors that could influence} \textit{not} developing or \textit{abandoning\slash stopping} the development of AI systems.
Then, we reviewed the titles and abstracts of the first 50 search results from each search source, and catalogued for further review those discussing formal mechanisms (e.g., exit plans, injunctions or suspensions) or more informal challenges (e.g., public backlash, data collection issues, etc.) that could influence pausing or stopping AI development. 
Through this process, we identified 48 initial sources. 
Then, using a snowballing approach \cite{wohlin_successful_2022} we explored relevant citations included within these 48 initial sources, gathering 32 additional relevant sources. 
{Since the goal of this scoping review was to identify broad, diverse factors contributing to AI abandonment, we did not systematically record information on works screened out as not relevant.

From the collected works that referenced formal and informal factors influencing AI development decisions, inductive thematic analysis \cite{braun_using_2006} was performed by the lead researcher to identify factors that could contribute to AI development challenges and abandonment. These factors were grouped into higher-order categories based on shared themes, which were iterated and discussed with the broader research team until consensus was reached.

After establishing the six broad categories of factors contributing to abandonment, we performed category-specific literature searches to address any gaps from the initial literature review and ensure relevant coverage. 
For example, after we established \textit{development lifecycle challenges} (see \S\ref{sec:lifecycle}) as a contributing category, we performed additional searches related to both `AI/ML development lifecycle' and its individual stages (e.g., `problem formulation,'). These category-specific searches increased the reviewed corpus by an additional 80 sources. 
The six categories identified via inductive thematic analysis and indicative literature review works underpinning each relevant factor are detailed in Table \ref{table:taxonomy}, with further contextual discussions on the individual factors now presented in \S\ref{sec:ethics}--\S\ref{sec:resources}.

\begin{table}[!hb]
\scriptsize
\setlength{\tabcolsep}{2pt}
\begin{tabularx}{\textwidth}{@{}>{\raggedright\arraybackslash}p{1.9cm} >{\raggedright\arraybackslash}p{2.2cm} >{\raggedright\arraybackslash}p{4.8cm} >{\raggedright\arraybackslash}p{2cm} >{\raggedright\arraybackslash}p{2.2cm} >{\raggedright\arraybackslash}p{2.4cm}@{}}
\toprule

\centering \textbf{Ethical Concerns (adapted from \cite{slattery_ai_2025})} & \centering \textbf{Stakeholder Feedback} & \centering \textbf{Development Lifecycle Challenges} & \centering \textbf{Organizational Dynamics} & \centering \textbf{Resource Constraints} &  {\centering \textbf{Legal\slash Regulatory Concerns}\par} \\ \midrule
Discrimination \newline
Toxicity \newline
Privacy \& security \newline
Misinformation \newline
Malicious actors \& misuse \newline
Concerns over overreliance \newline
Concerns over unsafe use \newline
Loss of human agency \& autonomy \newline
Labor displacement \newline
Environmental concerns \cite{hollanek_eu_2025} &
Stakeholder engagement \cite{kallina_stakeholder_2024} \textit{OR} \newline
Resistance \cite{devrio_building_2024} from:  \newline
$\bullet$ Employees \cite{abdalla_100000_2025, nedzhvetskaya_role_2024} \newline
$\bullet$ Users \cite{devrio_building_2024, jurca_integrating_2014} \newline
$\bullet$ Data subjects \cite{ajmani_secondary_2025} \newline
$\bullet$ Civil society \cite{data_center_watch_64_2025, stop_gen_ai_stop_2025} \newline
$\bullet$ Domain experts \cite{hogg_stakeholder_2023} \newline
$\bullet$ Activists \cite{ajmani_secondary_2025} &
Could not measure target variable \cite{kernahan_burying_2025, tal_target_2023, guerdan_measurement_2025} \newline
Challenging to collect desired data \cite{madaio_assessing_2022, ashmore_assuring_2022} \newline
Dataset too small \cite{berisha_digital_2021, cross_bias_2024, vabalas_machine_2019} \newline
Lacked ground truth \cite{guerdan_groundless_2023, guerdan_measurement_2025} \newline
Too challenging to pre-process/curate data \cite{sambasivan_everyone_2021, passi_trust_2018, werder_establishing_2022} \newline
Too challenging to label data \cite{bernhardt_active_2022, sylolypavan_impact_2023, zajac_ground_2023, gondimalla_aligning_2024} \newline
Inappropriate model selection \cite{lones_avoiding_2024, raji_fallacy_2022} \newline
Model training too technically challenging \cite{lones_avoiding_2024, cobbe_understanding_2023} \newline
Model training too resource intensive \cite{wu_sustainable_2021, dodge_measuring_2022} \newline
Undefined or inappropriate success\slash evaluation criteria \cite{lones_avoiding_2024, biderman_pitfalls_2021} \newline
Model performance not sufficient \cite{ahmed_ai_2024,raji_fallacy_2022} \newline
Difficult to integrate into pipelines \cite{badampudi_software_2016, russo_navigating_2024} \newline
Inability to conduct evaluation\slash pilot \cite{raji_fallacy_2022} \newline
Low adoption in pilot deployment \cite{estrada_mit_2025} &
Changing incentives/priorities \cite{ivanova_as_2025} \newline
Not indicated/scoped well \cite{lin_funding_2025, raji_fallacy_2022, black_toward_2023} \newline
Not aligned with organizational strategy \cite{rakova_where_2021} \newline
Insufficient leadership sponsorship \cite{marwaha_deploying_2022, lichtenberg_this_2025} \newline
Clients\slash customers did not want it \cite{stop_gen_ai_stop_2025, gartner_gartner_2024} \newline
Concerns over keeping company assets private \cite{cobbe_understanding_2023} &
Lacked technical expertise to build \cite{noauthor_quantifying_2021} \newline
Lacked technical expertise to maintain \cite{hogg_stakeholder_2023} \newline
Too expensive to build \cite{gartner_gartner_2024} \newline
Too expensive to maintain \cite{tobias_mann_most_2025} \newline
Compute availability concerns \cite{appenzeller_navigating_2023, burch_deep_2025} \newline
Compute cost concerns \cite{witzenberger_microsoft_2025, burch_deep_2025} \newline
Development timeline too long \cite{hogg_stakeholder_2023, kuutila_time_2020} \newline
Cheaper to outsource \cite{whang_contracting_1992, cobbe_understanding_2023, simon_escape_2024} \newline
Easier to outsource \cite{badampudi_software_2016, whang_contracting_1992, simon_escape_2024} &
AI regulations compliance \cite{hollanek_eu_2025, anderljung_frontier_2023, fitriyah_eus_2024} \newline
Data protection regulations \cite{pistilli_stronger_2023, chappidi_accountability_2025, norval_data_2019} \newline
Domain-specific regulations \cite{nazi_large_2024, rezaeikhonakdar_ai_2024} \newline
Deemed too high risk \cite{fitriyah_eus_2024} \newline
Lack of regulatory guidance \cite{pulkkinen_systemic_2025} \newline
Liability concerns \cite{shumway_medical_2024, terranova_ai_2024, saenz_autonomous_2023} \newline
Intellectual property concerns \cite{lucchi_chatgpt_2024, zhang_privacy_2024, amon_uncertain_2026} \\
\bottomrule
\end{tabularx}
\caption{Taxonomy of factors contributing to AI non-development or abandonment, with illustrative references from the reviewed literature corpus indicated.}
\label{table:taxonomy}
\vspace{-25pt}
\end{table}

\subsection{Non-Development Category 1: Ethical Concerns}

\label{sec:ethics}

Our analysis of factors mediating AI abandonment identified growing ethical concerns over the individual and/or societal risks of developing AI systems that could be grouped into three main categories.\footnote{Since our analysis is guided by the overarching question of factors motivating AI abandonment,
we discuss \textit{ethical concerns} in this paper from a perspective of that likely used in AI organizational practice, rather than through a more formal moral or philosophical lens.} First, there are concerns that AI deployments can bring negative societal impacts, including if they lead to job loss, or provide inconclusive evidence due to its probabilistic nature \cite{morley_what_2020}. 
Second, there are concerns over AI being used for ends themselves deemed unethical across various principles (e.g., weapons deployment) \cite{abdalla_100000_2025}.  
Finally, there are critiques that the process of constructing and implementing AI systems itself can perpetuate individual or societal harms, such as by extracting resources or labor from already marginalized communities \cite{gray_ghost_2019, mohamed_decolonial_2020} or enabling discrimination, often through the guise of `objectivity' via automation \cite{morley_what_2020}.

Many RAI frameworks have analyzed societal risks and related ethical issues within AI systems \cite{wong_seeing_2023, rismani_measuring_2025, tabassi_artificial_2023, weidinger_taxonomy_2022, hutiri_not_2024}, including AI risk repositories. 
\citet{slattery_ai_2025} perform a comprehensive meta-review of 777 AI risks identified across 43 separate taxonomies, synthesizing the reviewed taxonomies into 7 main risk domains. 
{To reflect common themes identified across AI risk taxonomies, these risk domains were adapted from \citet{slattery_ai_2025}} in \S\ref{sec:taxonomy} to include factors involving \textbf{Discrimination}, \textbf{Toxicity}, issues with \textbf{Privacy \& security} (also discussed in \S\ref{sec:legal}), \textbf{Misinformation}, \textbf{Malicious actors \& misuse}, \textbf{Concerns over overreliance}, \textbf{Concerns over unsafe use}, \textbf{Loss of human agency \& autonomy}, \textbf{Labor displacement}, and \textbf{Environmental concerns} (including the hardware and energy resources required to sustain development \cite{hollanek_eu_2025, wu_sustainable_2021, dodge_measuring_2022}).  
}

\subsection{Non-Development Category 2: Stakeholder Feedback}
Stakeholder feedback may affect decisions to abandon AI development when explicitly solicited through \S\ref{sec:stakeholder-engagement} \textbf{stakeholder engagement} performed by organizations, or when externally imposed through \S\ref{sec:resistance} \textbf{resistance and/or collective actions}. 
Note that feedback may often communicate information reflected in other categories of factors driving abandonment (e.g., concerns over ethics, low model performance, cost of development). As such, this section specifically considers the influence of the \textit{stakeholder feedback itself} (i.e., that the relevant concern is amplified or only taken seriously when levied through a stakeholder feedback mechanism). 
In these cases, {abandonment may be driven by key considerations including but not limited to}  reputational risks \cite{jeune_realharm_2025}, fear of repercussions, or general lack of public support \cite{johnson_fall_2024}.

\subsubsection{Stakeholder engagement.}
\label{sec:stakeholder-engagement}

The RAI community has advocated for stakeholder involvement throughout the development lifecycle to surface potential risks and harms, and shape AI system design to address actual stakeholder needs \cite{kallina_stakeholder_2024, kallina_stakeholder_2025, bell_think_2023, mushkani_right_2025, hickman_trustworthy_2021, corbett_power_2023,kallina_limits_2026}. 
Stakeholder engagement frameworks have been proposed across application domains, including education \cite{fuligni_would_2025}, medicine \cite{wang_stakeholder-centric_2025}, and public sector systems \cite{kawakami_situate_2024}, and may also be scoped by \cite{hickman_trustworthy_2021} or support \cite{bell_think_2023} regulatory guidance. 
Businesses also often rely on stakeholder engagement during project scoping and requirements gathering to ensure downstream success with users and clients, financial viability, and possibly avoid legal fees \cite{kallina_stakeholder_2024}. For example, 
\textit{agile} techniques advocate for frequent collaboration with \textbf{end-users} and \textbf{customers} to shape development processes \cite{kallina_stakeholder_2025, jurca_integrating_2014}. 
Through these processes, organizations may receive feedback from stakeholders to not build certain features or systems, or surface potential challenges to development that may lead to AI abandonment.

Many stakeholder engagement processes may only involve those directly tied to commercial interests \cite{kallina_stakeholder_2024, hickman_trustworthy_2021}; yet, other relevant stakeholders include \textbf{those impacted by the system} (e.g., communities subject to surveillance), \textbf{employees} building and maintaining systems, and \textbf{domain-specific actors} such as caregivers and administrators in the case of medicine \cite{hogg_stakeholder_2023}. \citet{nedzhvetskaya_role_2024} highlight the critical role of \textbf{workers} as stakeholders in AI development, arguing that while system ``designers'' are often given more prominence and power, ``trainers'' (including those generating, cleaning, and processing data) are often more likely to be subject to harm from such algorithms.
\citet{ajmani_secondary_2025} also advocate improved engagement with \textit{secondary stakeholders}, including \textbf{data contributors} and \textbf{activists}. 
At the same time, it can be challenging to identify cases where stakeholder involvement influences AI abandonment, as evidence is less likely to be published on systems that don't materalize.

\subsubsection{Resistance and/or collective action.}
\label{sec:resistance}

Stakeholder groups may also be mobilized outside of formal stakeholder engagement process, `disseminating' feedback via offline or online mechanisms \cite{johnson_fall_2024}  or engaging in explicit resistance. 
\textbf{Employees} tasked with developing AI systems can advocate abandonment of AI projects, with \citet{abdalla_100000_2025} analyzing technology worker resistance actions, 
frequently employed to halt AI development contracts over purposes found objectionable by employees (e.g., military uses).
Collective action by employees through organized labor \cite{freeman_two_1979} or by affected communities \cite{pi2026pushpushback} may also be used to shape and resist AI development practices \cite{nedzhvetskaya_role_2024, boag_tech_2022}.

Resistance can also be employed by other parties, including \textbf{data subjects} (those whose data is used to train such algorithms \cite{burgess_how_2024}) or those whose \textbf{jobs would be replaced or altered} by AI development \cite{samuel_is_2025}. 
For example, users may push back against development of AI systems through prompt injections \cite{agnew_data_2024}, jailbreaks \cite{yu_dont_2024, guo_hey_2024, shen_anything_2024}, opting-out, and other guides on avoiding AI \cite{stop_gen_ai_stop_2025, samuel_is_2025}. 
\citet{devrio_building_2024} taxonomize how people respond ``from below'' towards algorithmic harms, including refusing legitimate engagement with systems. 
Collective labor movements have also fought for protections against AI development by those potentially subject to labor replacement, including actors and writer's guilds \cite{anguiano_how_2023} and specialized professions \cite{samuel_is_2025, center_for_labor_and_a_just_economy_bargaining_2025}. 
While these forms of actual or proposed user resistance may appear to reflect low AI adoption, these collective actions have implications for AI \textit{development}, as they can prohibit, ban, or require oversight on upstream AI development \cite{anguiano_how_2023, chappidi_accountability_2025}.

\subsection{Non-Development Category 3: Development Lifecycle Challenges}
\label{sec:lifecycle}

Common challenges during AI development can be pinpointed to different stages of the AI development lifecycle. These issues may drive abandonment if the challenges cannot be overcome or ultimately create a nonfunctional AI system.  
As outlined in \S\ref{sec:taxonomy-methodology}, we performed broad searches into literature specifically on challenges appearing throughout the AI development lifecycle \cite{ashmore_assuring_2022, qian_orchestrating_2020, paleyes_challenges_2023, cobbe_reviewable_2021, suresh_framework_2021, lee_risk_2021}, including problem formulation; data collection, pre-processing, and curation; model building, training, and fine-tuning; and model performance evaluation and monitoring. 
Then, we performed targeted literature searches on the broad stages described in \S\ref{sec:problem-formulation}-\S\ref{sec:evaluation} (adapted from \cite{cobbe_reviewable_2021, paleyes_challenges_2023}) to identify additional factors that may prevent completion of the development lifecycle.

\subsubsection{Problem formulation.}
\label{sec:problem-formulation}
Problem formulation involves translating a real-world problem into a well-defined task, including selecting appropriate target variables, metrics, and evaluation criteria.
More specifically, problem formulation has been linked as an early AI development process shaping later, or downstream, stages of fairness assessment 
\cite{passi_problem_2019, jacobs_measurement_2021a, jacobs_measurement_2021b}, ground truth construction \cite{tal_target_2023, lakkaraju_selective_2017, muller_designing_2021}, and performance evaluation \cite{passi_making_2020, wang_against_2024}. 
Yet, \citet{mao_how_2019} report that ML practitioners often struggle to understand alternative ways to construct questions and hypotheses.
\citet{kernahan_burying_2025} investigate ML target variable construct invalidity,
finding that many algorithms are \textbf{unable to measure their target variable}, thus contributing to implementation failure. \textbf{Undefined or inappropriate success\slash evaluation criteria} is a challenge surfaced in later stages often involving problem formulation choices \cite{lones_avoiding_2024, biderman_pitfalls_2021}.

\subsubsection{Data curation.}
Data curation involves collecting, cleaning, labelling, and preprocessing data for use in training and evaluation of ML models. 
It may be \textbf{challenging to collect desired data} \cite{roh_survey_2021, biderman_pitfalls_2021}, leading to missing and imbalanced datasets that result in performance issues and bias in high-stakes settings, including clinical decision-making \cite{tasci_bias_2022, amal_use_2022, esteva_dermatologist-level_2017, flores_unsupervised_2021, lipkova_artificial_2022}. 
\textbf{Datasets that are too small} can result in overfitting \cite{berisha_digital_2021, cross_bias_2024, vabalas_machine_2019}, potentially leading organizations to abandon certain use cases in favor of other uses cases enabled by data they can access. 
Other studies of ML practitioners and general ML data practices have identified \textbf{challenges in pre-processing and curating data} \cite{wan_how_2020}, including opaque data processing habits and insufficient handling of data from minority protected attributes  \cite{sambasivan_everyone_2021, passi_trust_2018, werder_establishing_2022}.
It can also be \textbf{challenging to label data} in various domains, particularly as labor-intensive, subjective labeling processes may create label bias and low inter-rater reliability \cite{sylolypavan_impact_2023, bernhardt_active_2022}. \textbf{Lack of ground truth} target labels is also a frequent challenge  \cite{guerdan_groundless_2023, guerdan_measurement_2025, chappidi_manual_2025}, and can introduce measurement bias to systems when proxies for an actual yet unmeasurable outcome are used \cite{guerdan_groundless_2023, fogliato_fairness_2020, tal_target_2023, zajac_ground_2023, decamp_latent_2020}. 

\subsubsection{Model building}

\textbf{Inappropriate model selection} can influence AI functionality and abandonment of systems \cite{raji_fallacy_2022}. For example, models may be too simple, too complex, or involve incorrect assumptions for the available data \cite{lones_avoiding_2024}. Increasingly complex model architectures and/or algorithmic supply chains can also make \textbf{model training too technically challenging}. For example, deep learning architectures often require sophisticated interpretability techniques to investigate non-linear interactions \cite{lones_avoiding_2024}, and foundation models can require additional database/software dependencies that complicate AI system development \cite{cobbe_understanding_2023, lones_avoiding_2024, hopkins_ai_2025, widder_dislocated_2023}. Model training and inference can also be \textbf{resource intensive} lifecycle stages \cite{wu_sustainable_2021}, influencing decision-making where resources such as GPUs or data center storage are limited.

\subsubsection{Evaluating \& monitoring real-world performance.}
\label{sec:evaluation}
Many AI systems may be abandoned at the evaluation stage if \textbf{model performance is not sufficient}, including incorrect predictions \cite{ahmed_ai_2024}, hallucinations \cite{ashktorab_emerging_2025, burke_researchers_2024}, or unexpected system behavior \cite{raji_fallacy_2022}. Depending on prior system design decisions, evaluations may suffer from \textbf{undefined or inappropriate success\slash evaluation criteria} \cite{biderman_pitfalls_2021, lones_avoiding_2024}. For example, inappropriate evaluation metrics may prevent rigorous statistical evaluations \cite{biderman_pitfalls_2021} or provide misleading results (e.g., F1 score better captures performance over accuracy for highly imbalanced datasets) \cite{lones_avoiding_2024}. 
Pilot deployments may also influence abandonment \cite{estrada_mit_2025}, where \textbf{low adoption} indicates that the developed system will not be used. Lastly, \textbf{integration into existing workflows} is a key factor mediating success of AI system deployments \cite{badampudi_software_2016, russo_navigating_2024, greenhalgh_beyond_2017, marwaha_deploying_2022}.

\subsection{Non-Development Category 4: Legal\slash Regulatory Concerns}
\label{sec:legal}

Law and regulation, including those  principle-based \cite{floridi_unified_2019, correa_worldwide_2023, sun_principles_2025, pistilli_stronger_2023} or domain-specific, can directly impact whether system development proceeds. This section considers the challenges of complying with law and regulation and the consequences of non-compliance which may impact AI system non-development and abandonment.

\textbf{Concerns over complying with (emerging) AI regulation} may stop or slow down AI development \cite{fitriyah_eus_2024}, including instances of `regulatory flight' \cite{anderljung_frontier_2023} where organizations choose to not develop technologies in certain regions due to more stringent regulatory restrictions. 
Non-development may occur where systems are deemed \textbf{too high risk}, either explicitly prohibited if deemed to be of `unacceptable' risk under regulation such as the EU AI Act \cite{hollanek_eu_2025} or bringing additional `high-risk' regulatory compliance requirements that organizations may not wish to undertake \cite{fitriyah_eus_2024}. 
\textbf{Lack of regulatory guidance}, or uncertainty over how to interpret emerging regulations, may also slow or hamper development \cite{pulkkinen_systemic_2025, floridi_unified_2019, hudig_its_2026}.

Law and regulation concerning data rights, protection, and ownership can also shape decision-making surrounding AI development. \textbf{Data protection mechanisms} such as the EU's GDPR can impose restrictions on appropriate data collection, storage, and use \cite{pistilli_stronger_2023} that shape organizational decisions to pursue AI development \cite{norval_data_2019}. For example, \citet{chappidi_accountability_2025} find evidence that record-keeping mechanisms and requirements, including those mandated by GDPR, can influence and even dissuade practitioners from building systems that record data subject to such protections. 
Similarly, \textbf{domain-specific data privacy and protection regulations} can also shape decision-making to pursue AI development. For example, systems using individually identifiable health information (PHI) may be subject to additional regulatory requirements \cite{nazi_large_2024}, bringing additional compliance costs, considerations, and potential regulatory bodies/actions. 
Generative AI integrations have also driven scrutiny around data ownership rights, copyright law \cite{pistilli_stronger_2023, zhang_privacy_2024}, and  \textbf{intellectual property} \cite{lucchi_chatgpt_2024}. 
Compliance with privacy and copyright protection laws may include opt-out provisions \cite{amon_uncertain_2026}, machine unlearning \cite{amon_uncertain_2026, zhang_privacy_2024}, or system removal altogether. 
Abandonment over intellectual property concerns may also occur given ongoing lawsuits seeking injunctions on AI system sales over copyright infringements \cite{lucchi_chatgpt_2024}.

Lastly, legal concerns can also emerge outside specific internal and external regulatory frameworks, including through {fear of} \textbf{liability risks} and lawsuits \cite{mansi_implications_2025}. Organizations may seek to avoid risks of AI development involving product liability \cite{raji_fallacy_2022}, or stakeholders may hold domain-specific concerns over individual legal responsibility when AI systems are developed and used, including in medicine \cite{hogg_stakeholder_2023, harned_machine_2019, rowland_digital_2022, terranova_ai_2024, price_potential_2019}.

\subsection{Non-Development Category 5: Organizational Dynamics}
\label{sec:org-dynamics}

Organizational dynamics, including tensions between individuals, teams, and organizations and their own incentives, workflows, and mandates, can shape AI development practices and related decisions to abandon AI systems. 
Organizational incentives to pursue technology development \cite{stol_key_2014} often involve decisions on revenue-generating directions for the organization \cite{marwaha_deploying_2022}. \textbf{Leader sponsorship} is also a critical factor shaping outcomes of AI system development, with \citet{marwaha_deploying_2022} arguing the necessity of both an ``internal champion''  and ``executive sponsor.'' AI development at the organization-level has been driven by hopes of profitability, \cite{ibm_institute_for_business_value_2025_2025}, ``AI-first'' policies \cite{morris_going_2025}, and other incentives. 
At the same time, \textbf{changing priorities or incentives} can lead to abandonment of AI development strategies, such as retail company Klarna backtracking on its AI-first policy \cite{ivanova_as_2025}.
Increasing algorithmic supply chain complexity, including through growing AI integrations with foundation models, may bring concerns over \textbf{keeping {company, user, or client} information or assets private} due to limited visibility into other parts of the supply chain \cite{cobbe_understanding_2023}.

Misalignment or \textbf{lack of alignment between organizational strategies} and individual/team operations can affect the success of AI development. 
\citet{rakova_where_2021} study organizational RAI implementation, identifying ``misalignment between individual and team incentives and org-level mission statements.'' 
\citet{kawakami_studying_2024} also identify disconnects between frontline worker concerns and employees with higher institutional power across public sector agencies developing AI. Organizations may also struggle with faithful translation between goals, data, and computational problems \cite{passi_problem_2019, kallina_limits_2026}, resulting in projects \textbf{not well scoped/indicated} \cite{lin_funding_2025, raji_fallacy_2022, black_toward_2023}.

\subsection{Non-Development Category 6: Resource Constraints}
\label{sec:resources}

Resource constraints may also drive organizational decision-making to not develop or abandon development.
AI systems may be \textbf{too expensive to build}, bringing costs associated with the system itself (e.g., employee compensation, hardware, energy, etc. \cite{henshall_billion-dollar_2024}) and those related to mitigating risks (e.g., stakeholder engagement, legal\slash regulatory compliance). 
Comparing development costs to value generated by such systems, \citet{ibm_institute_for_business_value_2025_2025} find that only 25\% of CEOs currently report returns on investment in AI. 
Systems also carry additional \textbf{costs to maintain}, potentially evidenced by reports that only 16\% of AI initiatives have scaled across enterprises \cite{tobias_mann_most_2025}. 
\textbf{Compute cost} and \textbf{compute availability concerns} can also mediate decision-making, with reported shortages in GPUs critical for model training \cite{appenzeller_navigating_2023}, increasing demand for data centers \cite{data_center_watch_64_2025}, and energy bottlenecks in sustaining data centers \cite{burch_deep_2025}. Costs to maintain generative AI frontier models significantly outpace revenue currently generated by subscriptions, creating further concerns of costs increasingly being passed to users, including through price hikes \cite{witzenberger_microsoft_2025}.

Another constrained resource may include \textbf{technical expertise to build and maintain} AI systems. 
AI systems are growing in complexity due to increasing data needs, complex ML architectures, and dependencies introduced by other AI system components \cite{neumann_cascading_2025, hopkins_ai_2025, cobbe_understanding_2023}. Organizations may {lack necessary resources to maintain} AI systems \cite{hogg_stakeholder_2023}. 
For example, some tech executives report that AI development is scaling down (due to complexity) \cite{sauer_google_2024}, while 54\% of surveyed CEOs report hiring for AI-related roles that did not exist a year ago \cite{ibm_institute_for_business_value_2025_2025}.

AI abandonment may also be influenced by \textbf{development timelines that are too long} \cite{kuutila_time_2020, hogg_stakeholder_2023}.
Resource constraints relating to costs, time, and expertise may also drive abandonment of in-house development in favor of \textbf{outsourcing} development \cite{badampudi_software_2016, kuutila_time_2020, whang_contracting_1992} or contracting, purchasing, or procuring off-the-shelf products  \cite{badampudi_software_2016, johnson_legacy_2025, hudig_its_2026}.

\subsubsection*{Summary.}

Table \ref{table:taxonomy} establishes and summarizes the six categories of factors contributing to AI abandonment identified from inductive thematic analysis applied to our scoping literature review. 
{While existing work on AI nonuse or abandonment tends to focus on factors mediating technology \textit{adoption} and related user-focused models, our taxonomy presents an expanded lens to analyze AI abandonment \textit{throughout the 
development lifecycle}. As such, our taxonomy emphasizes both decisions made prior to system deployment and factors beyond user-level interactions, including broader organizational and related stakeholder dynamics.}

While categories in \S\ref{sec:taxonomy} are presented separately for clarity, these factors  are often linked/related across categories, and decisions to abandon AI development are likely shaped by multiple factors overlapping, interacting, and/or layering on top of each other. 
For example, stakeholder feedback may raise concerns over ethical risks, likely development lifecycle challenges, or reveal conflicting organizational dynamics. As such, our taxonomy enables the identification of multiple categories that layer and interact to eventually trigger abandonment of AI development.

\section{Analyzing Real-World Cases of AI Abandonment}
\label{sec:cases}

Next, we sought to gather evidence and analyze how the factors identified in \S\ref{sec:taxonomy} influenced decisions to abandon AI system development in real-world practice. 
\subsubsection*{Overall Methods} 
We gathered and analyzed case studies on abandoned AI systems from both AI incident repositories and an empirical AI practitioner survey. 
Data from the AI incident repository in \S\ref{sec:ai-incidents} enabled analysis of factors leading to abandonment in publicly documented systems (mostly reflecting those reaching deployment stages), while the practitioner survey in \S\ref{sec:survey} augmented our investigations with further data on systems that are abandoned at various stages, including pre-deployment.
Details on case filtering, data collection, and analyses are described in \S\ref{sec:incident-methods} and \S\ref{sec:survey-recruitment}-\S\ref{sec:survey-design}.
We present initial case details and summaries in \S\ref{sec:results}, and discuss broader trends and findings on real-world AI system abandonment in \S\ref{sec:discussion}. 

\subsection{Analyzing Abandoned AI Case Studies via AI Incident Repositories}
\label{sec:ai-incidents}
The AI, Algorithmic and Automation Incidents and Controversies (AIAAIC) Repository crowdsources documentation of algorithmic and/or AI-related risks, harms, and incidents \cite{aiaaic_ai_2025}. {Since the AIAAIC repository relies on publicly available details and reports, many entries originate from \textit{deployed} systems and frequently discuss ethical concerns.} We analyzed AIAAIC case studies to investigate factors contributing to \textit{why} AI systems are abandoned.

\subsubsection{Methods}
\label{sec:incident-methods}
The AIAAIC Repository web database (accessed 18/10/2025) was reviewed for curated data categories, including system details like the purpose, technology type, and developers/deployers, and impacts of the harm or controversy. 
We filtered for cases using the `Response Taxonomy' column, which describes actions taken \textit{by the individuals or organisations who developed/deployed the AI system}, including employee termination, public apology, policy review, and system suspension. 
We chose to filter on the `Response Taxonomy' field as it would most closely capture the perspectives and decision-making by the relevant party to abandon AI development. 
We selected all AIAAIC cases mentioning key terms of [`termination,' `suspension,' `pause,' `closure,' `scrapped,' `deletion,' `cancellation,' 'retraction,' `loss,' 'disabled,' `withdrawal,' `halted,' or `removal'] in the `Response' column to collect abandoned AI system cases.

Then, {the lead researcher} reviewed relevant case details summarized directly within the repository entry and links to external coverage of the incident. 
Cases were excluded from further analysis if they did not involve decisions to \textit{not develop} or \textit{stop developing} AI systems (e.g., merely updating the system, eventually reinstating the system, only removing problematic data from training datasets). 
Finally, we applied the taxonomy identified in \S\ref{sec:taxonomy} to categorize and analyze factors leading to AI abandonment. 
Details from these sources were used to identify and code 1) the stage at which the AI system was abandoned (e.g., project scoping, data curation) and 2) relevant factors contributing to AI abandonment under the taxonomy outlined in the \S\ref{sec:taxonomy}.

\subsection{Analyzing Abandoned AI Case Studies via Practitioner Survey}
\label{sec:survey}

Building on work conducted in \S\ref{sec:ai-incidents}, we sought to expand our analysis of abandoned AI systems across diverse domains and industries, past those specifically scoped as `incidents' and likely to be mature enough in development to meet news reporting thresholds. As such, we collected further abandoned AI case studies through an empirical online practitioner survey to increase our reach, diversity, and spread of analyzed abandoned AI cases. Practitioners were asked to report, detail, and analyze cases where they abandoned AI development.

\subsubsection{Study population and recruitment}
\label{sec:survey-recruitment}

The survey targeted individuals involved in commissioning, designing, developing, implementing, and/or monitoring AI systems, including software engineers, product/project managers, data scientists, and stakeholders in compliance, legal, and ethics roles. Participants needed to self-report at least one relevant role; otherwise, the survey was terminated early and their responses were excluded. 
Recruitment occurred via open calls through professional networks (including LinkedIn, Slack, emailing lists, and relevant workshops), consistent with methods in similar studies \cite{kallina_stakeholder_2025}. Participation was voluntary and anonymous, with human research subjects approval obtained through the department ethics committee. 

\subsubsection{Survey design}
\label{sec:survey-design}

We designed a practitioner
survey to collect case studies on instances where AI development projects were abandoned or continued, and to explore factors and processes influencing these decisions. 
The survey collected cases where an individual, team, and/or organization chose to not develop or stop developing an AI system for \textit{any} reason. 
{The survey also asked practitioners to report cases where their organization continued AI development, despite their assessment that the system should have been abandoned.}
Both of these survey sections gathered details about the AI system, including its purpose, relevant training data, user population, and the application domain. 
Participants reported the stage at which development was (or should have been) abandoned and indicated relevant factors towards abandonment using category-specific select-all-the-apply questions based on \S\ref{sec:taxonomy}.
The order of these category-specific questions was randomized to minimize ordering effects. 
Lastly, the survey collected details on the decision-making process contributing to (or preventing) abandonment, including who was involved, tools or methods used, and under what circumstances the AI system might be reconsidered for future development.  
Finally, participant demographics were collected, including role, years of experience, industry, and country (reported in the Appendix).

\subsection{Results} 
\label{sec:results}

This section presents descriptive analyses of our collected case studies to characterize the diverse categories of factors contributing to AI non-development and abandonment, with further qualitative insights presented in \S\ref{sec:discussion}. 

\subsubsection{AI Incident cases.}  Out of 2109 total entries (as of 10/2025), 143 AIAAIC cases included a \textit{suspension} key word (\S\ref{sec:incident-methods}) in the `Response' category. 52 AIAAIC cases were excluded from further analysis for 1) not involving AI systems specifically (e.g., incidents concerning dataset content, scams incorrectly claiming AI involvement) or 2) not involving an AI system reported as abandoned (e.g., system was updated and/or re-introduced).

We ultimately identified 91 AIAAIC cases involving AI abandonment for further analysis, ranging from 2016 to 2025. 85\% (n=77) of analyzed incidents were abandoned post-deployment, with 5\% abandoned at the project proposal stage and 5\% abandoned at the pilot deployment stage. 
\textit{Ethical concerns} were reported as contributing to abandonment for 81 cases, with 51\% (n=41) raising concerns over bias\slash discrimination and 51\% (n=41) involving privacy and security concerns such as potential for surveillance. \textit{Stakeholder feedback} was relevant towards abandonment for 48 cases, frequently involving public backlash on social media from affected users or data subjects; expert criticism from civil rights, privacy, and other advocacy groups; and investigative reporting from news outlets and think tanks. 
\textit{Legal\slash regulatory concerns} reportedly contributed to abandonment in 38 AIAAIC cases, 
with data protection regulations and domain-specific regulations (e.g., anti-discrimination laws in housing, hiring, etc.) shaping concerns over AI systems. 
\textit{Development lifecycle challenges} were identified in 26 cases, with 23 specifically reporting issues or challenges with model performance during pilot or formal deployment. 
\textit{Resource constraints} and \textit{organizational dynamics} also influenced abandonment in 4 and 3 AIAAIC cases, respectively. 

\begin{table}[!h]
\centering
\scriptsize
\setlength{\tabcolsep}{2pt}\renewcommand{\arraystretch}{1.2}\begin{tabularx}{\textwidth}{@{}
>{\raggedleft\arraybackslash}m{5cm}
>{\centering\arraybackslash}m{0.2cm}
>{\centering\arraybackslash}m{1cm}
>{\centering\arraybackslash}m{2.7cm}
>{\centering\arraybackslash}m{0.7cm}
>{\centering\arraybackslash}m{0.9cm}
>{\centering\arraybackslash}m{1.2cm}
>{\centering\arraybackslash}m{0.9cm}
>{\centering\arraybackslash}m{0.9cm}
>{\centering\arraybackslash}m{0.4cm}
>{\centering\arraybackslash}m{0.4cm}
@{}}

\textbf{AI System Description} & {\textbf{ID}} & {\textbf{Abandoned?}} & {\shortstack[c]{\textbf{Stage} \\[-2pt] \textbf{Relevant to} \\[-2pt] \textbf{Abandonment}}} & \rotatebox{55}{\shortstack[c]{\textbf{\scriptsize Ethical} \\[-2pt] \textbf{\scriptsize Concerns}}} & \rotatebox{55}{\shortstack[c]{\textbf{\scriptsize Organizational} \\[-2pt] \textbf{\scriptsize Dynamics}}} & \rotatebox{55}{\shortstack[c]{\textbf{\scriptsize Development} \\[-2pt] \textbf{\scriptsize Lifecycle} \\[-2pt] \textbf{\scriptsize Challenges}}} & \rotatebox{55}{\shortstack[c]{\textbf{\scriptsize Stakeholder} \\[-2pt] \textbf{\scriptsize Feedback}}} & \rotatebox{55}{\shortstack[c]{\textbf{\scriptsize Resource} \\[-2pt] \textbf{\scriptsize Constraints}}} & \rotatebox{55}{\shortstack[c]{\textbf{\scriptsize Legal/Regulatory} \\[-2pt] \textbf{\scriptsize Concerns}}} \\ 

\midrule
{\setlength{\baselineskip}{2em}\textbf{\tiny Chatbot for public assistance program applications}} & P1 & Yes & Deployment & $\checkmark$ & $\checkmark$ & $\checkmark$ & & $\checkmark$ & \\
{\setlength{\baselineskip}{0.8em}\textbf{\tiny AI-mediated e-commerce integration w/ web search}} & P2 & Yes &  Model retraining & & $\checkmark$ & $\checkmark$ & & $\checkmark$ & \\
{\setlength{\baselineskip}{0.8em}\textbf{\tiny Clustering to identify fraudulent crypto users}} & P3 & Yes &  Model retraining & & & $\checkmark$ & & & \\
{\setlength{\baselineskip}{0.8em}\textbf{\tiny Clustering to identify high-risk labor operations}} & P4 & Yes & Problem operationalization & $\checkmark$ & & $\checkmark$ & $\checkmark$ & & \\
{\setlength{\baselineskip}{0.8em}\textbf{\tiny Inference to improve data quality gaps}} & P5 & Yes & Data collection & & $\checkmark$ & $\checkmark$ & & & \\
{\setlength{\baselineskip}{0.8em}\textbf{\tiny In-house LLM development (to compete w\slash industry)}} & P6 & Yes & Model evaluation & & $\checkmark$ & $\checkmark$ & & $\checkmark$ & \\
{\setlength{\baselineskip}{0.8em}\textbf{\tiny LLM-based infrastructure support for SWEs}} & P7 & Yes & Problem formulation & $\checkmark$ & $\checkmark$ & $\checkmark$ & $\checkmark$ & $\checkmark$ & $\checkmark$ \\
{\setlength{\baselineskip}{0.8em}\textbf{\tiny Concept extraction from larger NLP model}} & P8 & Yes &  Problem operationalization & & & $\checkmark$ & & $\checkmark$ & \\
{\setlength{\baselineskip}{0.8em}\textbf{\tiny LLM-based notification summaries for SWEs}} & P9 & Yes &  Problem operationalization & & $\checkmark$ & $\checkmark$ & $\checkmark$ & $\checkmark$ & \\
{\setlength{\baselineskip}{0.8em}\textbf{\tiny RAG system for sustainability business contacts}} & P10 & Yes & Data collection & & $\checkmark$ & & $\checkmark$ &  & \\
{\setlength{\baselineskip}{0.8em}\textbf{\tiny Decision trees for COVID-19 travel warnings}} & P11 & Yes & Model evaluation & & $\checkmark$ & $\checkmark$ & $\checkmark$ & & \\

{\setlength{\baselineskip}{0.8em}\textbf{\tiny LLMs for functional form identification}} & P12 & Yes & Model evaluation & & & $\checkmark$ & $\checkmark$ & & \\

{\setlength{\baselineskip}{0.8em}\textbf{\tiny LLM summarization of  congressional\slash policy press}} & P13 & No &  During development & & & $\checkmark$ & & & \\

{\setlength{\baselineskip}{0.8em}\textbf{\tiny Predict system bugs using troubleshooting logs}} & P14 & No &  During development & & & & & $\checkmark$ & \\

{\setlength{\baselineskip}{0.8em}\textbf{\tiny LLM support to improve vulnerability management}} & P15 & No &  During development & & $\checkmark$ & $\checkmark$ & $\checkmark$ & $\checkmark$ & \\

{\setlength{\baselineskip}{0.8em}\textbf{\tiny Extract lease details from unstructured documents}} & P16 & No &  Before development & & &  & $\checkmark$ & & \\

{\setlength{\baselineskip}{0.8em}\textbf{\tiny Public health outbreak detection}} & P17 & No &  During development & & & $\checkmark$ & & $\checkmark$ & \\

\bottomrule
\end{tabularx}
\caption{{Surveyed case studies indicate a wide array of factors contributing to abandonment, with many reporting abandonment prior to deployment.} P13-P17 reflect cases where participants reported that the system should have been abandoned due to the indicated factors, but the organization continued with development. Factor-level details contributing to cases indicating abandonment are reported in \ref{fig:case-level-factors}. SWEs = software engineers}

\vspace{-30pt}
\label{table:case-study-factors}
\end{table}

\subsubsection{Survey cases.}
A total of 36 complete survey responses were collected, with 8 responses removed from participants who did not report involvement with design, development, or deployment of AI systems. Of the remaining 28 practitioner responses, we collected 12 cases of abandoned AI system development, 5 cases of AI systems that participants reported should have been abandoned {in their view} (but development was pushed forward regardless), and 23 cases of AI system development reported as justifiable or suitable. Participants reported a wide range of factors contributing to AI abandonment, including most frequently \textit{development lifecycle challenges}, followed by \textit{resource constraints}, \textit{organizational dynamics}, and \textit{stakeholder feedback}. Factors contributing to abandonment are detailed by category in Table \ref{table:case-study-factors} and individually in Figure \ref{fig:case-level-factors}.

\subsubsection*{Summary.}
A majority of AIAAIC cases analyzed indicated \textit{ethical concerns} and related \textit{stakeholder feedback} as key factors contributing to AI abandonment (Table \ref{table:incident-factors}). This finding is expected since these cases were gathered via a crowdsourced AI incident database cataloging public sources more likely to describe mature, deployed AI systems.
Less than 5\% of all AIAAIC repository entries involved AI abandonment, indicating that non-development is not a frequently documented system outcome. We also identified other non-ethics-related factors driving abandonment across these cases, including \textit{legal\slash regulatory concerns} and \textit{development lifecycle challenges}.

In contrast, cases collected via survey tended to report non-ethics concerns as influencing abandonment (see Tables \ref{table:case-study-factors} and \ref{table:incident-factors}). 
Surveyed cases also usually reported multiple categories of factors as contributing to abandonment, most frequently including \textit{organizational dynamics}, \textit{development lifecycle challenges}, and \textit{resource constraints}. While 85\% of AIAAIC cases were abandoned post-deployment, many of the surveyed abandoned cases reported issues \textit{prior} to deployment, including during problem operationalization and model adjustment. As such, our initial analysis of abandoned AI cases via survey indicate a greater influence of internal factors in deciding to abandon AI development, particularly where systems do not make it to the deployment stage.

\begin{table*}[!htbp]
\centering
\small
\setlength{\tabcolsep}{2pt}\renewcommand{\arraystretch}{1}\begin{tabularx}{\textwidth}{@{}>{\centering\arraybackslash}m{3cm}|>{\centering\arraybackslash}m{1.8cm}
>{\centering\arraybackslash}m{2.1cm}
>{\centering\arraybackslash}m{2.1cm}
>{\centering\arraybackslash}m{2.1cm}
>{\centering\arraybackslash}m{2.1cm}
>{\centering\arraybackslash}m{1.8cm}
>{\centering\arraybackslash}m{2.1cm}
@{}}
\toprule
\textbf{Factors Contributing to Abandonment} & \textbf{Ethical Concerns} & \textbf{Organizational Dynamics} & \textbf{Development Lifecycle Challenges} & \textbf{Stakeholder Feedback} & \textbf{Resource Constraints} & \textbf{Legal\slash Regulatory Concerns} \\ 
\midrule
\textbf{\# of AIAAIC cases (n=91)} & \cellcolor{blue!69}81 & \cellcolor{blue!7}10 & \cellcolor{blue!15}23 & \cellcolor{blue!39}45 & \cellcolor{blue!4}4 & \cellcolor{blue!31}37 \\
\hline
\textbf{\# of survey cases (n=17)} & \cellcolor{blue!19}3 & \cellcolor{blue!46}9 & \cellcolor{blue!63}14 & \cellcolor{blue!37}8 & \cellcolor{blue!46}9 & \cellcolor{blue!6}1 \\
\bottomrule
\end{tabularx}
\caption{Our analysis of AI incident repository cases naturally indicates that ethical concerns, stakeholder feedback, and legal\slash regulatory concerns are primary influences on AI abandonment. Meanwhile, surveyed cases indicating abandonment frequently report factors beyond those 3 categories. Darker shaded cells reflect a higher proportion of cases in the category.} 
\label{table:incident-factors}
\vspace{-20pt}

\end{table*}

\section{Discussion}
\label{sec:discussion}

We now discuss and synthesize findings from our analyses of abandoned AI case studies gathered via AI incident repository and practitioner surveys. We also highlight gaps and opportunities for the RAI community to consider, address, and engage with AI abandonment and non-development as practices influencing RAI development and governance. Practitioner survey IDs [P\#] assigned in Table \ref{table:case-study-factors} and AIAAIC-generated source ID codes [AIAAIC\#] providing evidence are indicated in brackets where relevant.

\subsubsection*{Ethical concerns are not the only factors that contribute to abandoned AI development.}
Many abandoned systems analyzed from the AI incident repository identified concerns over potential or realized harm, including discrimination, toxicity, etc. This finding is not surprising given that these repositories seek to categorize instances of harm created by algorithms, and reflects the utility of these resources in establishing real-world risks and harms created by AI system development. 
At the same time, cases collected via practitioner survey frequently reported non-ethics reasons across resource constraints, organizational dynamics, and challenges during the AI development lifecycle as key factors in deciding to abandon AI development. 
While it remains critical to address ethical concerns when deciding whether to pursue AI development, we identify and evidence diverse, frequently overlapping drivers or levers influencing decision-making to abandon AI development. These levers can indicate different avenues to motivate change within organizations, and the relative influence and relationship between these levers should be further explored and investigated in RAI work.

\subsubsection*{Few surveyed participants reported using formal processes to determine whether to even proceed with building AI} 
While other works discuss public responses and influences on decision-making to abandon harmful algorithms \cite{johnson_fall_2024, devrio_building_2024}, our surveyed cases often involved abandoned AI systems that did not explicitly flag ethical concerns or receive negative stakeholder feedback. As such, these scenarios often required \textit{internal} reflections on whether to abandon AI development. Most participants reported no specific individual, team, or organizational decision-making processes, including P7 specifically flagging the need for ``a centralized AI strategy for the organization.''

Some RAI works advocate for formal processes to rule-in AI use, including problem formulation \cite{raji_fallacy_2022} and exploration of alternative methods \cite{lones_avoiding_2024}. Yet, very few RAI artifacts explicitly advocate or address AI abandonment as an available, acceptable, or appropriate decision {throughout the development process}. This is echoed by \citet{kawakami_situate_2024} finding that ``most existing toolkits assume that the decision to develop a particular AI system has already been made'' and \citet{wong_seeing_2023} discussing `solutionism' present in RAI toolkits that do not suggest ``fundamental changes to the corporate values systems or business models that may lead
to harms from AI system.''
Resources that reference not commissioning or de-commissioning systems \cite{tabassi_artificial_2023} often do not highlight or suggest conditions where abandonment may be appropriate. As such, our findings emphasize the need for processes that guide organizations in reflecting on conditions indicating or requiring AI abandonment.

\subsubsection*{Resource constraints can underpin many additional factors driving abandonment, bringing nuance to decision-making behind non-development.}

Many surveyed case studies reported \textit{resource constraint}s as a key factor in deciding to abandon AI development. 
These survey responses often indicated that procurement of existing tools was cheaper and easier, 
and that they lacked sufficient compute resources or dataset sizes [P6] to effectively maintain and justify their AI system.
Resource constraints may often drive factors indicating abandonment in other categories of the taxonomy. For example, \textit{insufficient dataset sizes} within development lifecycle challenges may be driven by insufficient funds or technical expertise to collect, curate, and appropriately pre-process data. In the case of La Buona Scuola teachers’ mobility algorithm in Italy [AIAAIC0705], a \textit{lack of technical expertise to build} the system (translated as ``a bombastic, redundant and non-maintenance-oriented system'' by \citet{lopera_algorithm_2020} from an Italian audit report) resulted in an \textit{inability to conduct an evaluation} of the system, ultimately driving abandonment.

Costs may be an additional (and potentially more salient) lever that encourages non-development or abandonment for organizations, even for AI systems with identified ethical concerns. For example, in the case of Alpaca [AIAAIC1029], a language model criticized for ethical concerns over \textit{misinformation}, a Stanford spokesperson only discussed in their public statement that \textit{costs to maintain} factored into the abandoned AI development \cite{quach_stanford_2023, landymore_stanford_2023}. Other categories of abandonment factors may also carry their own associated costs that influence overall decision-making to pursue AI development. For example, emphasizing the costs of lawsuits or regulatory probes into data protection compliance issues could make abandonment a more salient option for AI systems with ethical \textit{privacy \& security} risks. Similarly, emphasizing the costs associated with systems abandoned post-deployment due to \textit{stakeholder resistance} can make the upfront costs of \textit{stakeholder engagement }\cite{kallina_stakeholder_2024, kawakami_responsible_2024} more reasonable.

\subsubsection*{Early-stage challenges in AI development lifecycle often influence abandonment.}

We identified two cases where organizations could not \textit{measure their target variable} indicating the critical influence of early-stage decision-making. P4 abandoned their AI system at the problem operationalization stage, noting that ``although full ground truth is challenging to obtain, it is possible to collect offending vessels from news reports, or use IUU (illegal, unreported and unregulated) fishing list as a proxy.'' In contrast, the criminality prediction algorithm reported in AIAAIC0467 was abandoned after development, indicating potential challenges in identifying that the AI problem design was not possible \cite{raji_fallacy_2022}. Moreover, two surveyed case studies reported \textit{challenges in labeling data} [P3, P5] and four reported \textit{challenges in obtaining ground truth} [P3, P4, P5, P12]; yet, both P3 and P12 reported that development was abandoned at model training stages. This indicates that they may have proceeded past the data curation stage despite challenges, or that they only identified early-stage issues at the evaluation stage (e.g., P12 reporting ``we realized that the method our colleague was set on [...] was not scientifically rigorous or of potential interest to our broader research community.'') 
Thus, our work indicates the potentially beneficial role of deliberating on early-stage development challenges to identify abandoned AI projects sooner.

\subsubsection*{Resurrection of abandoned AI systems is a likely outcome.}
Many reviewed AIAAIC cases were excluded from final inclusion because the system was not actually abandoned, or in some cases ``reincarnated'' (as termed by \citet{johnson_fall_2024}). For example, reviewed AI systems were often suspended in one locality (e.g., Uber suspended their vehicles in Tempe, but then continued operations in Pittsburgh [AIAAIC0187]), for a certain period of time (e.g., Malta stopped their `Safe City' video surveillance program, but then appointed a new board to oversee the initiative years later [AIAAIC1051]), or re-introduced after a feature was tweaked (e.g., HireVue removed facial analysis from its screening algorithm [AIAAIC0579]). {These findings reflect how projects abandoned in the present can still have impacts (or `imprints' \cite{ehsan_algorithmic_2022}) in the future.}
Many surveyed cases also reported that their organization was likely to revisit development of abandoned systems, including under conditions involving ``better employee bandwidth'' [P1], ``more time'' [P2], and ``if we have close partnership with front line organisations'' [P4]. 
Surveyed cases also indicated that their organization realized it was cheaper or easier to outsource development [P6, P8, P9]. As such, our analyses echo the growing influence and relevance of algorithmic supply chains \cite{cobbe_understanding_2023}, AI-as-a-Service contracts \cite{cobbe_artificial_2021} (e.g., P16 reporting that they warned of issues over hallucinations but their client ``favored an AI-integrated solution first''), and procurement in RAI development \cite{johnson_legacy_2025, hudig_its_2026}.

\subsection{Summary \& Recommendations}

Our taxonomy developed in \S\ref{sec:taxonomy} and analysis of real-world abandoned AI development cases in \S\ref{sec:cases} indicate a diverse set of factors or levers influencing AI non-development. 
We call for further attention and reflection on AI abandonment and non-development as a multi-faceted practice, with the following recommendations for responsible AI communities:

\begin{tcolorbox}[sharp corners=south, rounded corners=north, colback=white!10!white, title=Recommendations]
\small \setlength{\parskip}{0pt}

\begin{enumerate}[left=0pt, label={\textbf{\arabic*.}}, itemsep=0.3ex, parsep=0pt]

\item {Responsible AI communities should \textbf{investigate, interrogate, and reflect on cases of AI abandonment occuring prior to deployment} to better understand current practices, challenges, and gaps in AI system development.} By documenting and analyzing cases of AI abandonment at all stages of development, the community can gain a more complete picture of drivers and blockers towards responsible AI development, including those non-ethics-related. 
\item {\textbf{Artifacts and tooling should explicitly increase visibility and outline discussions on non-development or abandonment}, particularly where organizations identify that they lack appropriate resources, expertise, or capacity to develop their desired system.} 
\item {Responsible AI development toolkits, frameworks, and artifacts can \textbf{make more explicit the influence of resources, including costs, technical expertise, and development timelines, on the potential success} of other development stages.} For example, encouraging organizations to explicitly outline the costs of collecting sufficiently large datasets or maintaining deployed systems can use resource constraints as a lever to discourage development of AI systems that also carry ethical risks or would face development lifecycle challenges. 
\item {Responsible AI tooling 
should explicitly \textbf{encourage organizations to reflect on, outline, and update conditions required to appropriately re-visit} or resurrect development of abandoned AI systems.}

\end{enumerate}
\end{tcolorbox}

\normalsize

\section{Limitations \& Future Work}

The taxonomy presented in \S\ref{sec:taxonomy} is meant to illustrate the diverse levers that can facilitate non-development rather than be exhaustive, and additional factors may emerge in future work.
While we aimed to review all available details, analysis of abandoned AIAAIC cases may not capture every single factor that ultimately contributed to AI abandonment, as these databases largely concern deployed systems and external reports may lack visibility into organizational dynamics or resource constraints. 
Relatedly, data on non-development and abandoned projects is hard to collect as these systems receive fewer resources, are often less formally documented, and can be less salient to practitioners. 
Our real-world data analyses affirmed this, as very few systems in the AIAAIC repository were reported as abandoned and surveyed practitioners were more inclined to report on continued AI developments rather than abandoned ones. 
Thus, our work {reflects a first step towards this gap}
by engaging and gathering insights from organizations that abandoned AI development across diverse domains and purposes, particularly in cases prior to system deployment. Future work can continue this approach, encouraging researchers, organizations, and advocacy groups to 1) continue collection, analysis, and sharing of abandoned AI cases across all lifecycle stages, 2) conduct further empirical analysis of incentives and drivers \textit{preventing} abandonment, {and 3) examine the broader impacts of AI system removal.}

\section{Concluding Remarks}
Our work seeks to increase visibility into AI non-development, departing from and complementing other works by conceptualizing abandonment as a practice enabled by diverse levers present throughout the AI development lifecycle. 
While academic and civil society resources frequently emphasize ethical risks as reasons to not develop AI systems, we find that decisions to abandon development often involve factors beyond ethical concerns, including development lifecycle challenges, organizational dynamics, and resource constraints. 
{Through this work, we indicate real-world incentives and levers that influence organizations to pursue or abandon AI development, and in turn, highlight actionable opportunities to better align RAI research and interventions with the factors that actually influence organizational decision-making.}
We urge the RAI community to further investigate AI abandonment, particularly including those abandoned in early-stages or prior to deployment, and advocate for updates to ongoing research and tooling
to support appropriate (dis)engagement with AI development.

\subsection*{Generative AI Usage Statement}
Asta (allen.ai) was used to identify links to potentially relevant academic sources, with any resulting academic sources reviewed directly in the externally linked/original source location (e.g., full PDF upload). Limited editorial support was provided from LLMs to adjust formatting commands for LaTeX tables (Copilot integration in VSCode, GPT-5 mini) and support ideation on rephrasing select statements (GPT-4o \& 5 mini).

\begin{acks}
We would like to thank the practitioner survey participants for their responses, as well as Anna Neumann and Anna Ida Hudig for their feedback on study materials.
SC is a PhD student in the NIH Oxford-Cambridge Scholars Program. This research was supported [in part] by the Intramural Research Program of the National Institutes of Health (NIH). The contributions of the NIH author(s) are considered Works of the United States Government. The findings and conclusions presented in this paper are those of the author(s) and do not necessarily reflect the views of the NIH or the U.S. Department of Health and Human Services.
\end{acks}

\bibliographystyle{ACM-Reference-Format}
\bibliography{references}

\renewcommand{\thefigure}{A\arabic{figure}}

\setcounter{figure}{0}

\renewcommand{\thetable}{A\arabic{table}}

\setcounter{table}{0}

\newpage
\appendix

\section{Abandoned AI Cases Reported Via Survey}
\subsection{Participant Demographics}
Participants represented 7 countries globally, with the majority from English-speaking nations. Responses were collected from the United States (n=14), the United Kingdom of Great Britain and Northern Ireland (n=6), India (n=3), Netherlands (3), Germany (n=3), and Spain (n=1). Participants reported an average of 12 years of professional experience (min=1 year, Q1=5, median=6, Q3=20, max=30 years). 

The surveyed cases reported operations across a wide range of domains, including technology (n=12), finance (9), research (9), government (6), professional services (n=5), health (3), energy (2), civil society\slash NGOs (2), military\slash defense (1), and legal (1). Participants also operated their algorithms in diverse industries, including tech companies (n=14), academia (n=8 with 6 related to AI research, 2 related to other fields), government (6), consulting firms (6), non-profits (n=6, 3 related to research and advocacy and 3 service-based), start-ups (3), and non-tech private/commercial companies (n=3).

\subsection{Factor-Level Details Contributing to Abandoned AI Cases}

\begin{figure*}[h]
    \centering
    \includegraphics[width=1\linewidth]{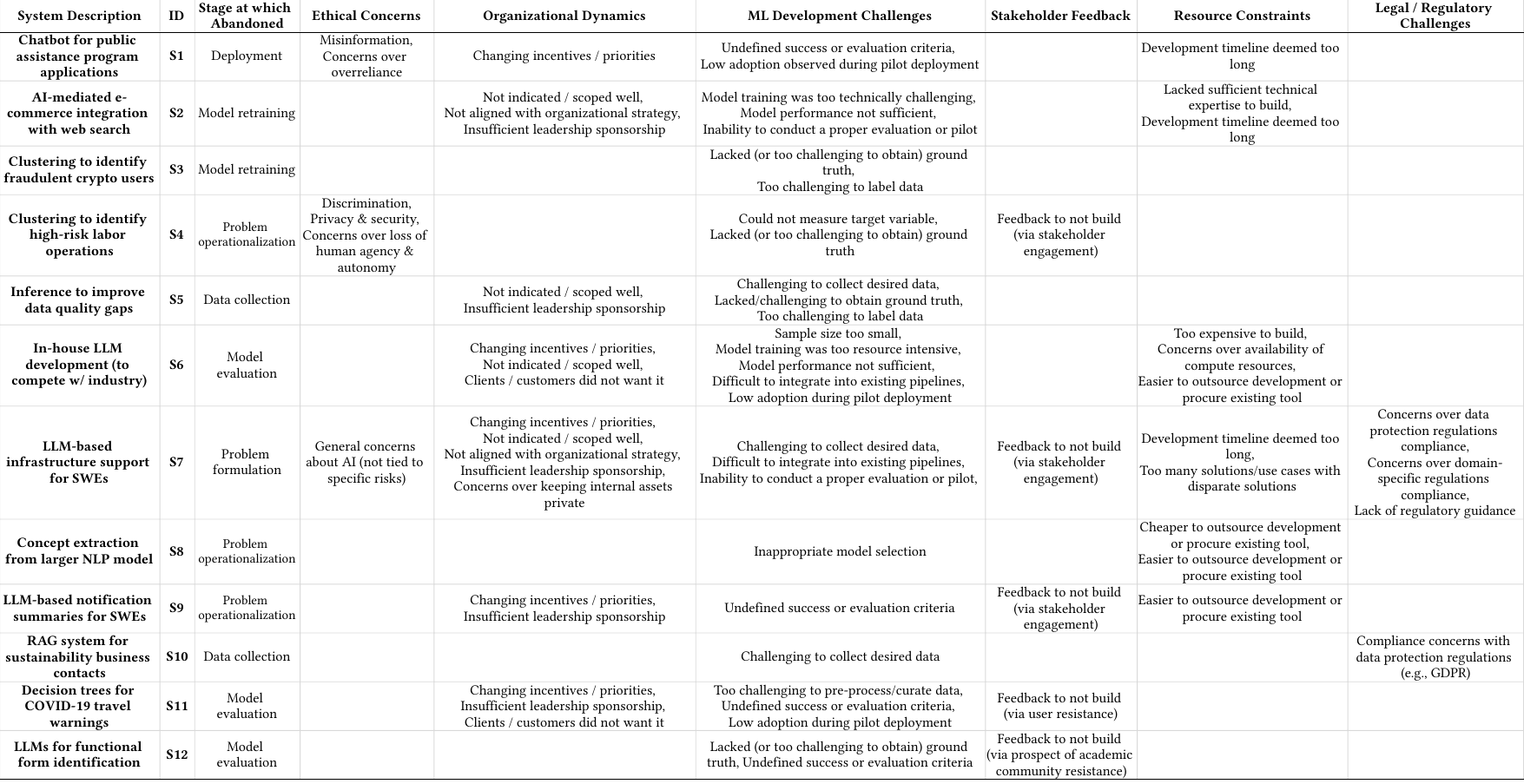}
    \caption{Factor-level details reported as contributing to AI abandonment decisions collected via practitioner survey.}
    \label{fig:case-level-factors}
\end{figure*}

\end{document}